\documentclass[twocolumn,11pt]{article}
\usepackage{lscape}
\usepackage{epsfig}
\usepackage{amssymb}
\usepackage{txfonts}
\usepackage{graphicx}
\usepackage{natbib} 

\begin{document}

   \title{ Interstellar H${\bf_2}$ toward HD 147888 }  

   \author{ {\bf Gnaci\'nski Piotr }  \\ 
   Institute of Theoretical Physics and Astrophysics,  \\
              University of Gda\'nsk,
              ul. Wita Stwosza 57, 80-952 Gda\'nsk, 
   }

   \date{Received \today / Accepted \today }

   \maketitle  
   
   \abstract{    
      The ultraviolet and far--ultraviolet spectra of HD 147888 allows to access the $\nu$=0 as well as higher vibrational levels of the ground H$_2$ electronic level. We have determined column densities of the H$_2$ molecule on vibrational levels $\nu$=0--5 and rotational levels J=0--6.
  The ortho to para H$_2$ ratio for the excited vibrational states equals to 1.2. For the lowest vibrational state $\nu=0$ and rotational level J=1 the ortho to para H$_2$ ratio is only 0.16. The temperature of ortho--para thermodynamical equilibrium is T$_{OP}=43\pm3$ K.
  
  The large number of H$_2$ absorption lines in the HST spectra allows to determine column densities even from a noisy spectra. 
  The measurements of H$_2$ column densities on excited vibrational levels (from the HST spectra) leads to constrains of radiation field in photon--dominated regions (PDR) models of interstellar cloud towards HD 147888.

   }   
   
   

\section{Introduction}

 The molecular hydrogen is the most common molecule in the interstellar medium. 
First detection of H$_2$ on vibrationally excited levels was performed by \cite{Federman} in HST (Hubble Space Telescope) spectrum of $\zeta$ Oph. 
A rich spectrum of vibrationally excited hydrogen molecule was described by \cite{Meyer} in the HST STIS (Space Telescope Imaging Spectrograph) spectrum of HD 37903. They have also noticed that vibrationally excited H$_2$ is present in the cloud towards HD 147888.

The presence of both FUSE (Far Ultraviolet Spectroscopic Explorer) and HST STIS spectra for HD 147888 ($\rho$ Oph D) gives us a rare opportunity to measure column density of H$_2$ energy levels excited by fluorescence ($\nu>0$) and also by collisions ($\nu$=0). In a consequence we can calculate a PDR model which is well constrained by density and by the ultraviolet radiation flux (star--cloud distance).

 \begin{figure*}
    \centering
    \includegraphics[width=\textwidth,viewport=1 1 840 500,clip]{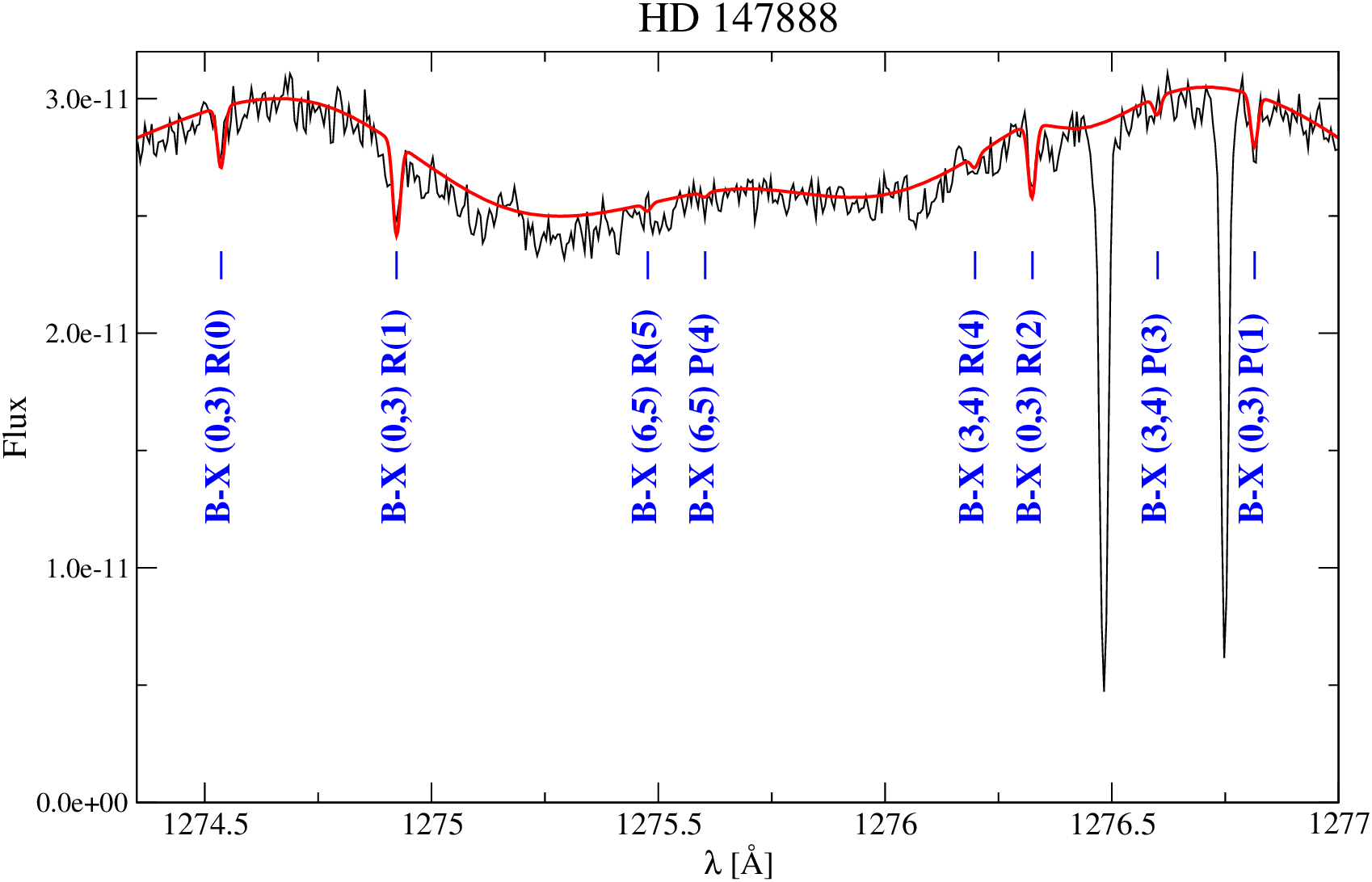}
    \caption{ A fragment of the HST spectrum. The synthetic spectrum (red line) was fitted to the whole HST spectrum. The Doppler broadening parameter was set to b$=2.5$ km/s. The two deep absorption lines are CI lines.     }
    \label{HST}
 \end{figure*}

 \begin{figure*}
    \centering
    \includegraphics[width=\textwidth,viewport=1 1 840 500,clip]{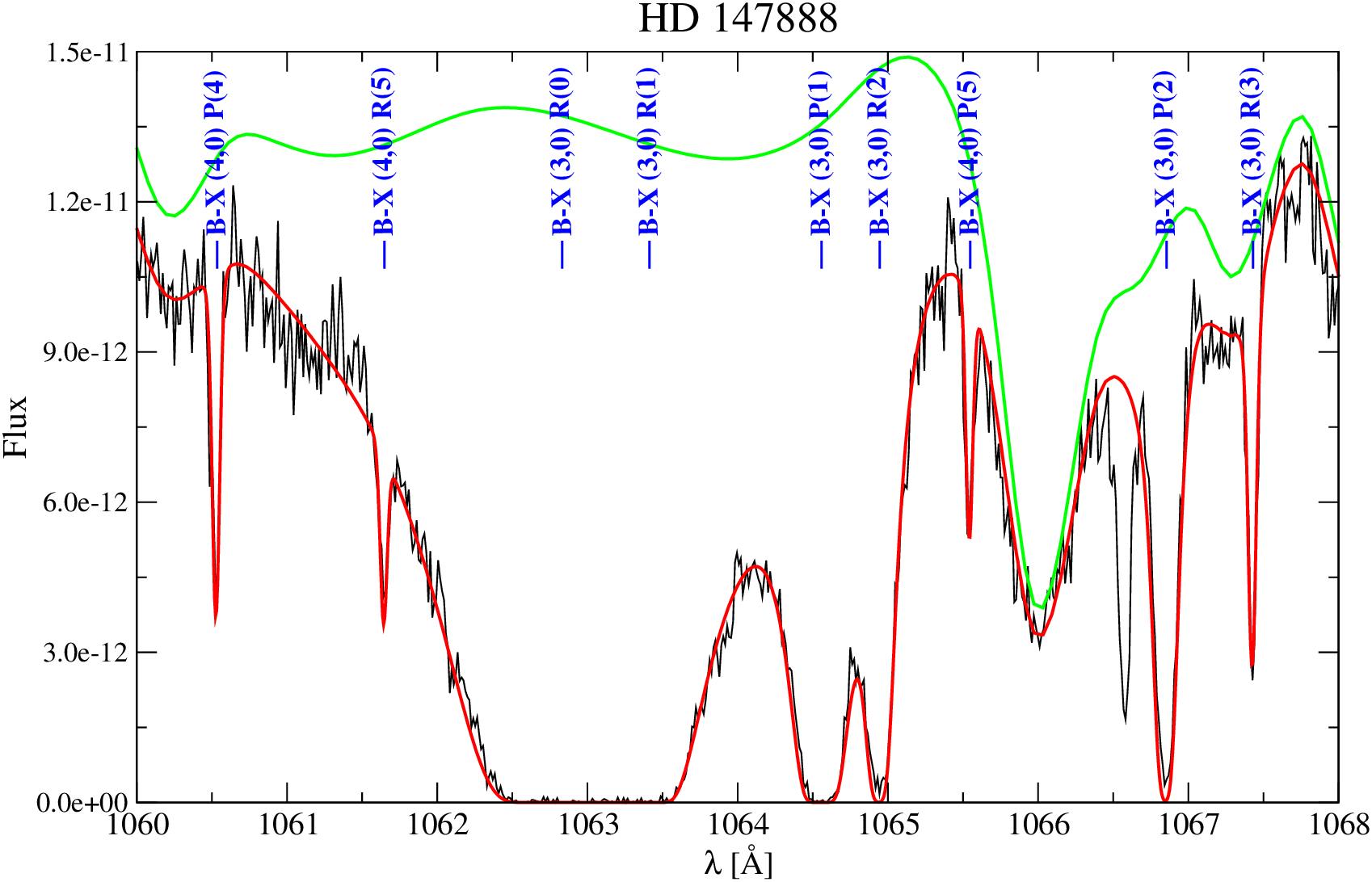}
    \caption{ A fragment of FUSE spectra with the B--X (3,0) transitions of H$_2$. The green line represents continuum and the red one is the simulated spectrum fitted to the observed one (black line). The absorption line at 1066.66 \AA \ is an Ar I line.     }
    \label{fuse}
 \end{figure*}

 \begin{figure*}
    \centering
    \includegraphics[width=\textwidth,viewport=1 1 680 480,clip]{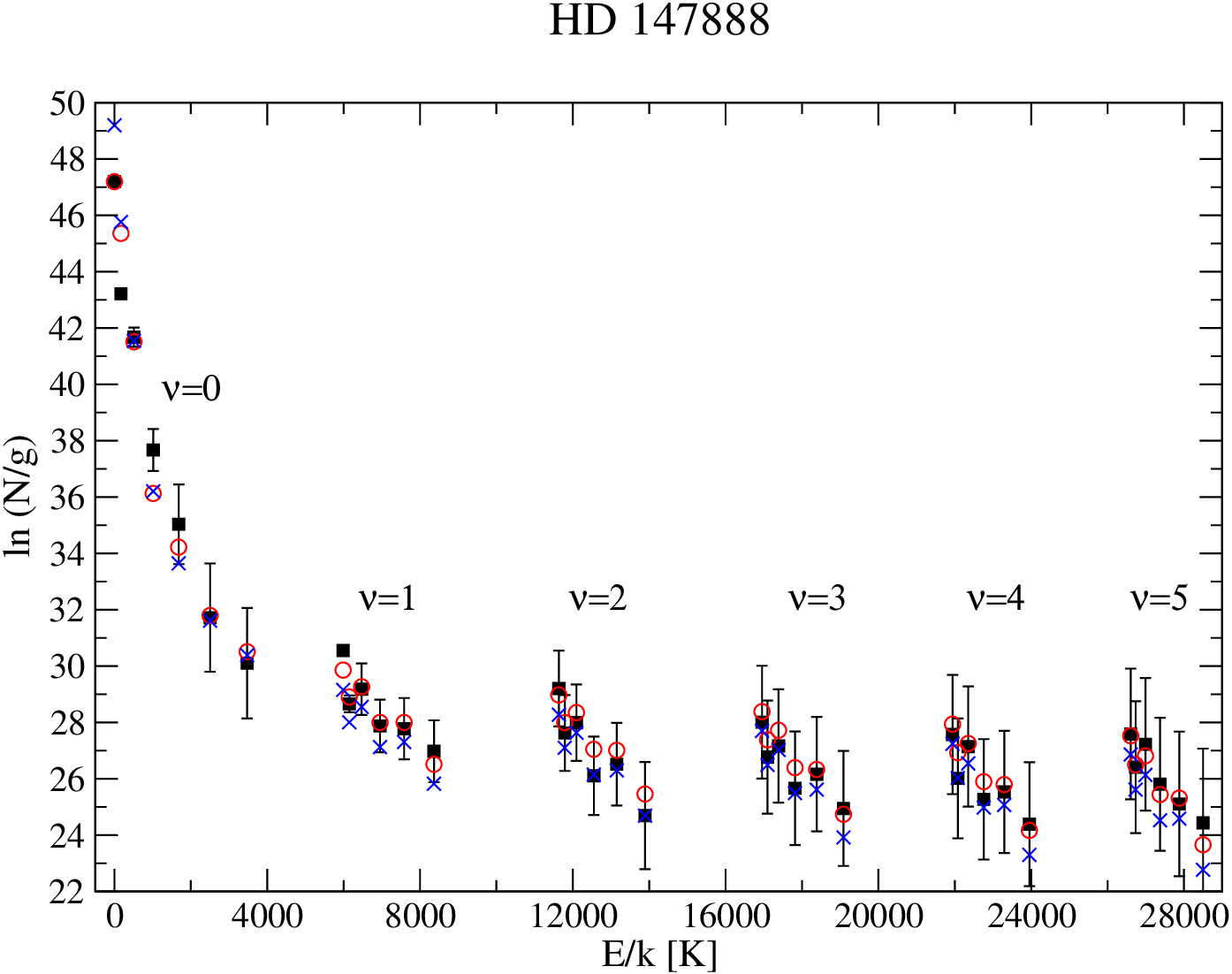}
    \caption{ Natural logarithms of H$_2$ column densities divided by statistical weight plotted versus energy (squares with error bar). The $\nu$ symbol denotes the vibrational quantum number. For $\nu$=0 rotational quantum numbers are J=0--6, and for vibrationally excited levels ($\nu>$0) J=0--5. The first PDR cloud model is shown as red circles. The second model is shown as blue crosses.   }
    \label{lnN-E}
 \end{figure*}

\section{Observations}

  We have used HST STIS spectrum o59s05010 to obtain column densities on vibrationally excited H$_2$ levels.
The STIS spectrum consists of 2 subexposures which were combined into a single spectrum ranging from 1160 \AA\ to 1356.8 \AA. The column densities of the ground vibrational levels ($\nu$"=0) were determined from the FUSE spectra P1161501016-19. These spectra were averaged using the IRAF tasks {\it poffsets} and {\it specalign}. 
The FUSE spectrum extends from 904 \AA\ to 1188 \AA.
We have used FUSE spectra from the following detectors 1blif4, 2alif4 and 1alif4 which have the best quality.

The whole HST STIS spectrum was fitted at once with a synthetic spectrum consisting of 355 H$_2$ absorption lines.
The cloud velocity and column densities of all vibrationally excited levels were free parameters.  The H$_2$ spectral line positions and oscillator strengths were adopted from \cite{Abgrall}. A fragment of HST spectrum together with a fit to the whole HST spectrum is shown on fig. \ref{HST}.

Each transition from ground $\nu$=0 J=0,1 levels to vibrationally excited levels of the B electronic level were fitted independently in the FUSE spectrum. 
We have calculated column densities from the following transitions: (0,0), (1,0), (2,0), (3,0) and (4,0). The first digit in parentheses denotes the upper vibrational level of the B electronic state. The second digit (always 0) is the vibrational level of the ground electronic level X.
At each point of the synthetic spectrum optical depth of spectral lines lying no more than 30 \AA\ were summed. The point spread function (PSF) for the FUSE spectra was a Gauss function with FWHM=15 km/s (\cite{Jensen}). The natural line widths were given by \cite{Abgrall-Atotal}. A synthetic spectrum together with the FUSE observation is shown on  fig. \ref{fuse}.
  The derived molecular hydrogen column densities are presented in table \ref{CD} and shown on fig. \ref{lnN-E}.
 
\section{Noise robustness}
 
 \begin{figure}
    \centering
    \includegraphics[width=\columnwidth,viewport=1 1 700 490,clip]{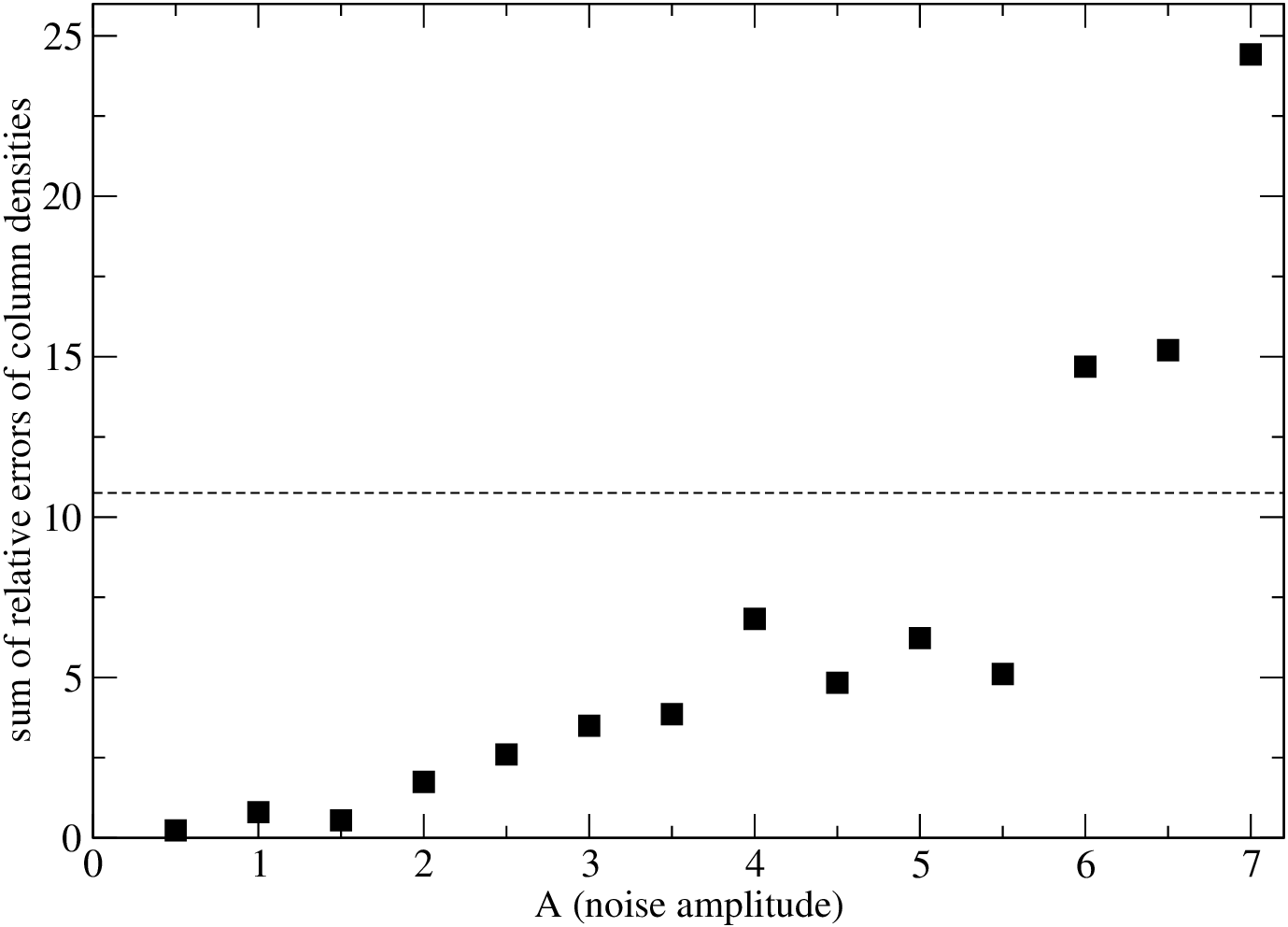}
    \caption{ Robustness of the synthetic spectrum fit to the noisy data.
       The dashed line shows errors introduced by changing the original column densities by 60\% (close to
       the quoted errors of column densities). }
    \label{Noise}
 \end{figure}
 
  Some of the H$_2$ absorption lines in the HST spectrum are at the noise level
(eg. absorption lines from the $\nu=1$ level). Therefore we have tested the 
robustness of the fit against noise. The noise observed on the HST STIS spectrum
is about 1\% at 1360 \AA\ and 3\% at 1160 \AA. We have added Gaussian noise 
of amplitude $A$ to the best fit synthetic spectrum. The points of the
generated noisy spectrum are at the same wavelengths as the HST spectrum 
used previously in the fitting procedure. The Gaussian noise has standard deviation $\sigma=1$.
The spectrum with noise was generated using the formula 
$F_A=F_{fit}(1+A(-0.01\lambda+14.6)\cdot0.01\cdot GaussianNoise)$, where the factor
$(-0.01\lambda+14.6)\%$ reflects the changing noise amplitude in the original HST spectrum.
The amplitude $A=1$ reflects the noise level in the original HST spectrum.
The $F_{fit}$ is a synthetic spectrum that represents the best fit to the observed HST STIS spectra.
  
  Next we have performed a fit to the generated spectrum $F_A$ and
obtained new column densities by fitting a new synthetic spectrum to the noisy one. All fits to the
noisy spectrum were performed with Doppler broadening parameter fixed to $b=2.5$ km/s. The
cloud velocity and all column densities for $\nu\ge1$ were free parameters. The continuum level
was common for all fits.
In order to evaluate the error introduced by noise of various magnitude $A$ we have
calculated a sum of squared relative errors:
\begin{equation}
\sum_{\nu\ge1,J=0-5} {\left(\frac{N_A(\nu,J))-N_0(\nu,J)}{N_{0}(\nu,J)}\right)^2}.
\end{equation}
The symbol $N_A$ represents column density calculated from spectrum with noise amplitude $A$.
The results are shown on fig. \ref{Noise}.
Even for the noise amplitude 5.5 times larger then in the HST spectra the
resulting column densities errors are significantly less to the 60 \% errors in the
best fit column densities. Errors about 60 \% are common for H$_2$ column densities derived
from the HST spectra (see tab. \ref{CD}).
So the fitting procedure is very robust against the noise.
Note however, that if the original HST spectrum had the noise 5 times larger, then
the continuum placement would be problematic or even impossible.
 
\begin{table*}
\centering
\caption{Column densities of the H$_2$ ro-vibrational levels towards HD 147888 [log cm$^{-2}$]. The absorption lines connected with the column densities marked in bold were clearly seen in the spectrum. Other lines are at the noise level. }
\label{CD}
  \scriptsize
\begin{tabular}{lcccccc}
\hline
J$\backslash\nu$ & 0 & 1 & 2 & 3 & 4 & 5  \\
\hline
0 & {\bf 20.50} $\pm$ 0.04 & 13.27 $\pm$ 0.05 & {\bf 12.68} $\pm$ 0.58 & {\bf 12.16} $\pm$ 0.87 & {\bf 11.97} $\pm$ 0.92 & {\bf 11.98} $\pm$ 1.01\\
1 & {\bf 19.72} $\pm$ 0.08 & 13.40 $\pm$ 0.13 & {\bf 12.95} $\pm$ 0.58 & {\bf 12.58} $\pm$ 0.87 & {\bf 12.25} $\pm$ 0.92 & {\bf 12.43} $\pm$ 1.02\\
2 & {\bf 18.80} $\pm$ 0.15 & 13.37 $\pm$ 0.40 & {\bf 12.86} $\pm$ 0.59 & {\bf 12.50} $\pm$ 0.87 & {\bf 12.49} $\pm$ 0.93 & 12.52 $\pm$ 1.02\\
3 & {\bf 17.68} $\pm$ 0.32 & 13.43 $\pm$ 0.40 & {\bf 12.66} $\pm$ 0.60 & {\bf 12.47} $\pm$ 0.88 & {\bf 12.30} $\pm$ 0.93 & 12.53 $\pm$ 1.03\\
4 & {\bf 16.17} $\pm$ 0.61 & 13.02 $\pm$ 0.47 &      12.47  $\pm$ 0.64 & {\bf 12.32} $\pm$ 0.88 & {\bf 12.04} $\pm$ 0.94 & 11.86 $\pm$ 1.12\\
5 & {\bf 15.29} $\pm$ 0.83 & 13.24 $\pm$ 0.48 &      12.24  $\pm$ 0.83 & {\bf 12.35} $\pm$ 0.89 &      12.11  $\pm$ 0.95 & 12.13 $\pm$ 1.14\\
6 & {\bf 14.19} $\pm$ 0.85 & -- & -- & -- & -- & --\\
\hline
\end{tabular}
\end{table*}

\section{Temperatures}
  
 \begin{figure}
    \centering
    \includegraphics[width=\columnwidth,viewport=1 1 700 490,clip]{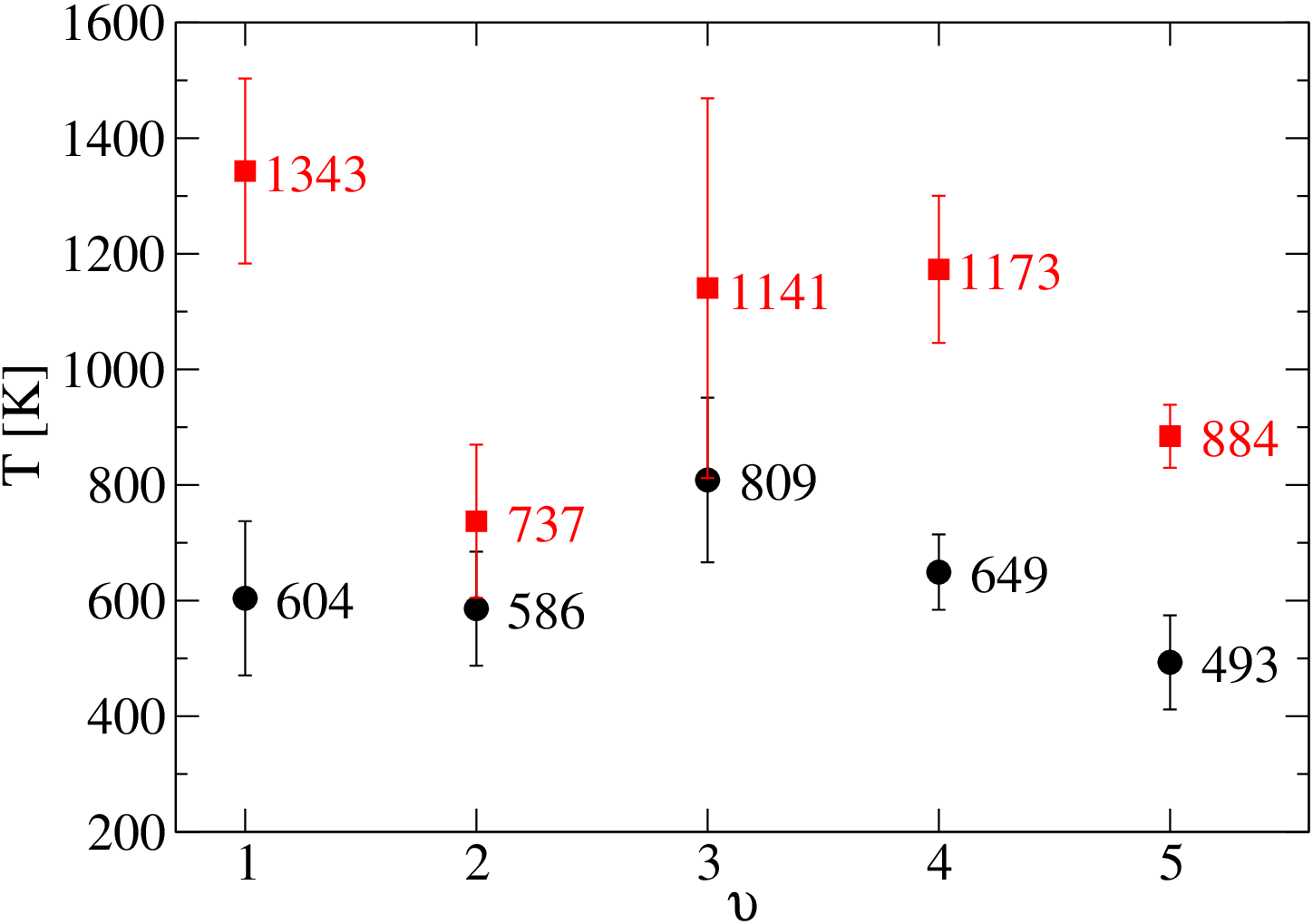}
    \caption{ The rotational temperature on vibrational levels 1--5. Red squares presents rotational temperature of ortho H$_2$, while the black circles are for para--H$_2$. }
    \label{Trot}
 \end{figure}

 The total observed column density of molecular hydrogen is equal to N(H$_2$)=$(3.74\pm0.36)\cdot10^{20}$.
 The column density of atomic hydrogen N(HI)=$(5.7\pm1.1)\cdot10^{21}$  cm$^{-2}$ was determined by \citet{Cartledge}. The resulting molecular fraction of hydrogen $f(H_2)=2N(H_2)/(N(HI)+2N(H_2))=0.11\pm0.02$. The ortho--para H$_2$ ratio for vibrational levels $\nu\ge$1 is O/P=$1.23\pm0.21$.  However the ortho--para H$_2$ ratio on the $\nu$=0 J=1 level is significantly lower. It was calculated using the method presented by \citet{Wilgenbus} and is equal to 0.16$\stackrel{+0.04}{\scriptstyle -0.02}$. 
 
The temperature of ortho--para H$_2$ equilibrium T$_{OP}$ was derived from the equation:
 \begin{equation} \label{eq_T01}
 \frac{N(1)}{N(0)}=\frac{ g(1)}{ g(0)} \exp{\left(-\frac{E(1)-E(0)}{kT_{OP}}\right)},
 \end{equation}
where N(J) are the column densities, E(J) is the energy on the J level ($\nu$=0) and $g(J)$ is the statistical weight
\begin{equation}
   \label{g}
   g(J)=\left\{
   \begin{array}{ll}
   (2J+1)  & \mbox{for para H$_2$ (even J)} \\
   3(2J+1) & \mbox{for ortho H$_2$ (odd J)}. \\ 
   \end{array}
   \right. 
 \end{equation} 
  The temperature of ortho--para H$_2$ equilibrium T$_{OP}=43\pm3$ K, similar to the value T$_{OP}=45\pm3$ K derived by \citet{Jensen}. 
 
 The T$_{02}$ rotational temperature involves only para--H$_2$ levels and was calculated from equation analogous to eq. \ref{eq_T01}.
The T$_{02}$ rotational temperature equals to $92\pm7$ K (\citet{Jensen} gives T$_{02}=88\pm1$ K).


The rotational temperature was also calculated for each vibrational level.
The rotational temperature is the inverse of line inclination taken with the minus sign on a plot: $ln(N/g)$ versus $E/k$ (see fig. \ref{lnN-E}). The line inclination was calculated with the linear regression method separately for ortho and para H$_2$ spin isomers. The resulting temperatures are presented on fig. \ref{Trot}. The vibrational level $\nu$=0 is populated by both collisions and fluorescence and cannot be fitted with a single temperature. 

For each excited vibrational level the points on the $ln(N/g)$ versus $E/k$ plot (fig. \ref{lnN-E}) are perfectly placed on a straight line (separately ortho and para states). The correlation coefficient is
between 0.92 and 0.996. The agreement with the Boltzmann distribution confirms the correctness of the
H$_2$ column densities.

\section{Models}

\begin{figure}
    \centering
    \includegraphics[width=\columnwidth,viewport=1 1 760 490,clip]{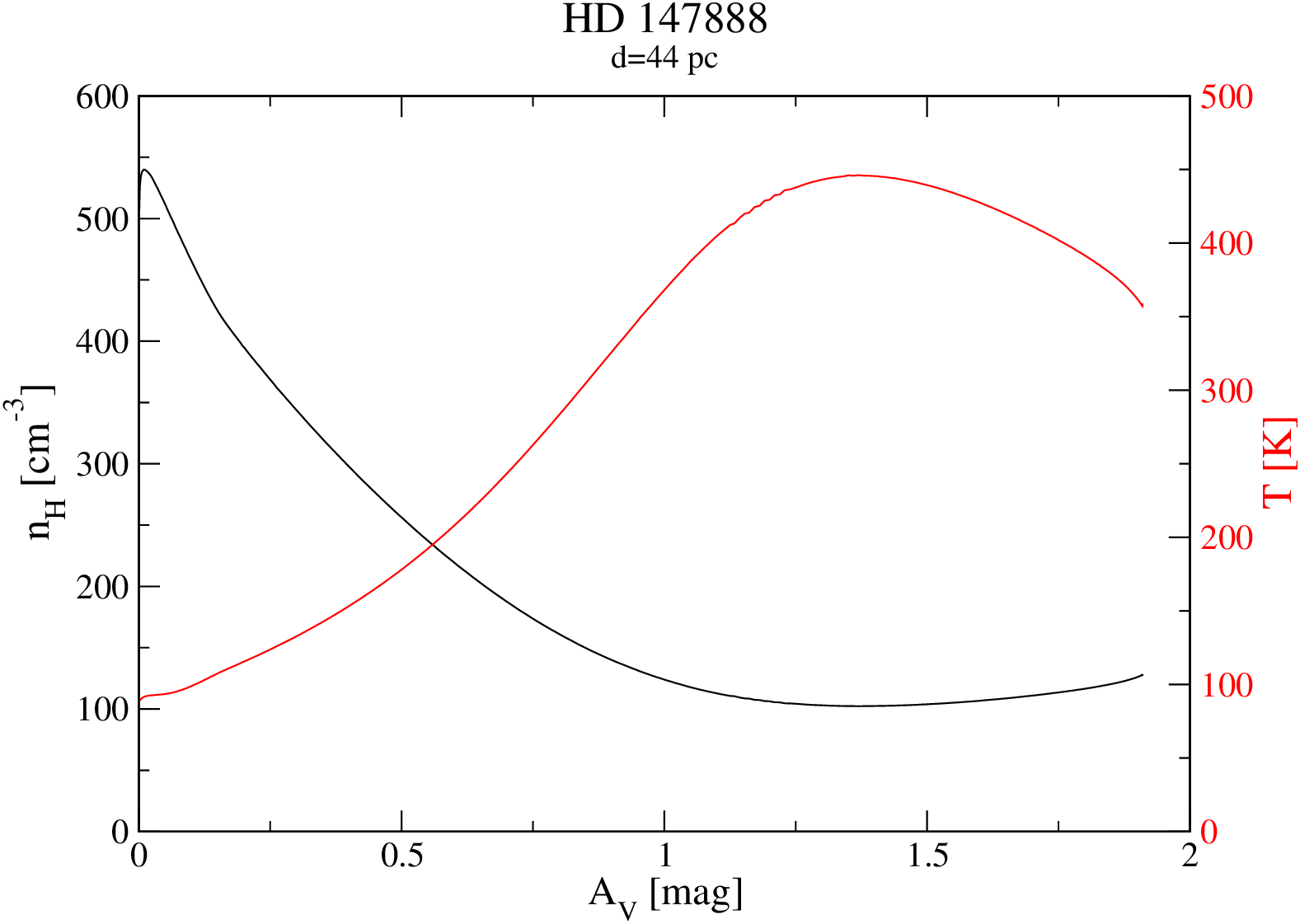}
    \caption{ The hydrogen density and kinetic temperature in the first cloud model (at d=0.44 pc). }
    \label{model44}
 \end{figure}
 
 \begin{figure}
    \centering
    \includegraphics[width=\columnwidth,viewport=1 1 760 490,clip]{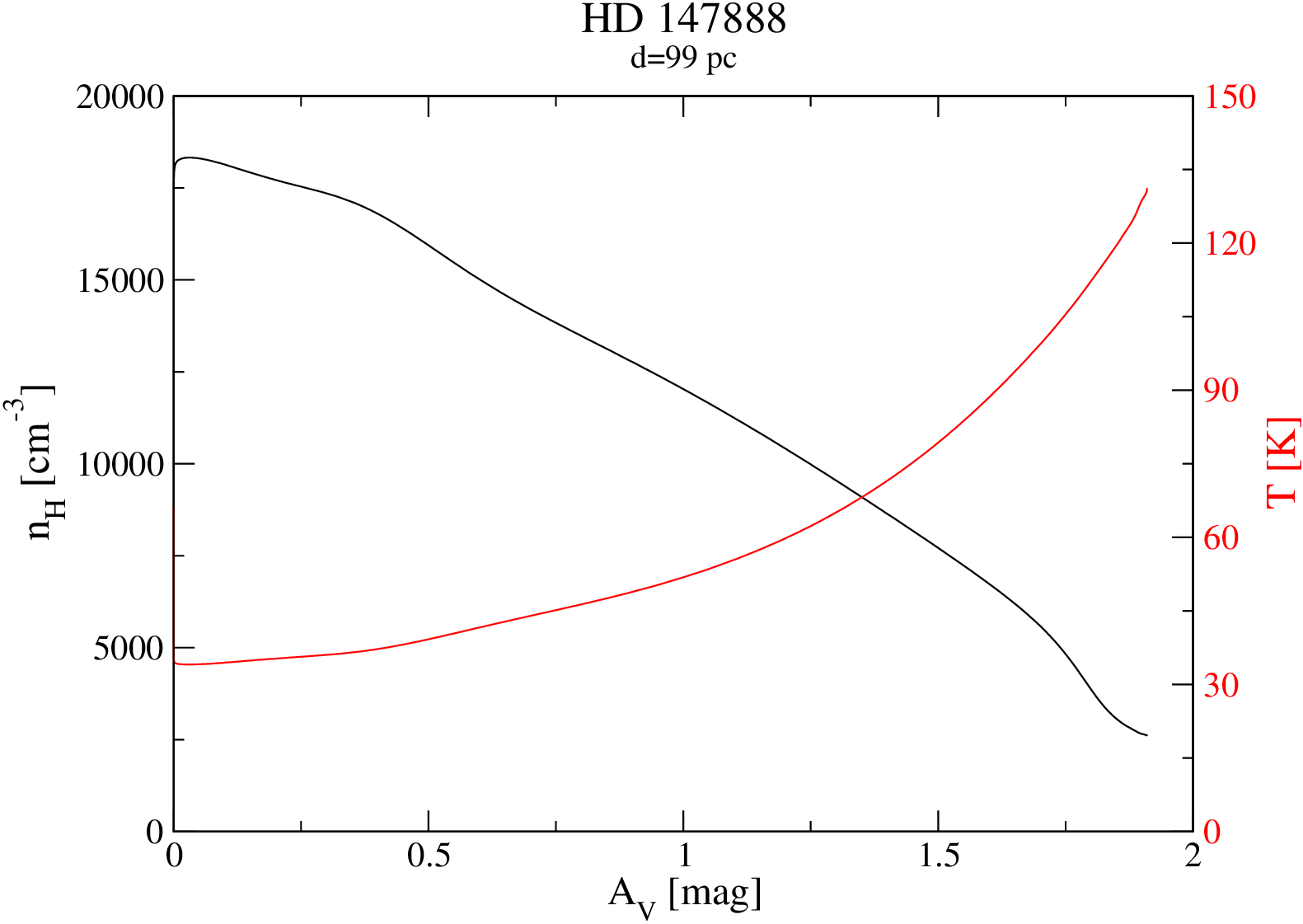}
    \caption{ The hydrogen density and kinetic temperature in the second cloud model (at d=0.99 pc). }
    \label{model99}
 \end{figure}
 
  We have used the Meudon PDR code \citep{Petit} to calculate the model of interstellar cloud in front of HD 147888. We have calculated isobaric models with exact radiative transfer in all H$_2$ spectral lines.
The interstellar radiation field on the observer side was always 1 Draine unit.
The turbulent velocity was set to $v_{turb}$=2 km/s and the star spectral type to B3V. 
The interstellar reddening E(B-V)=0.47, R$_V$=4.06 and A$_V$=1.91 were adopted from \citet{Rachford}.

The first PDR model was chosen among 1290 models by minimizing the difference of relative H$_2$ abundances observed in the cloud towards HD 147888 and calculated in a PDR model:
\begin{equation}
  \sum_{\nu,J} w_{\nu}\left(\log \frac{N_{obs}(\nu, J)}{N_{obs}(0, 0)}-\log \frac{N_{model}(\nu,J)}{N_{model}(0,0)}\right)^2
\end{equation}
The weights $w_\nu$ were chosen so, that the levels usually excited by fluorescence ($\nu>0$) have the some influence on the final sum, as the $\nu=0$ level excited collisionally.
The weights $w_\nu$ were equal to 5 for $\nu$=0 and 1 otherwise.

The first PDR model has a star--cloud distance of 0.44 pc. The interstellar radiation field on the star side is equal to 500 Draine units. The hydrogen density varies from 127 cm$^{-3}$ on the star side to 539 cm$^{-3}$ near the observers side. The gas kinetic temperature changes from 359 K on the star side to 90 K on the observer's cloud side. In this cloud model the ortho/para H$_2$ ratio for vibrationally excited states equals to 1.5.

Our second model was selected by minimizing the difference between the observed and calculated column densities:
\begin{equation}
  \sum_{\nu,J} w_{\nu}\left(\log {N_{obs}(\nu, J)}-\log {N_{model}(\nu,J)}\right)^2
\end{equation}
A model chosen using the above formula has to reproduce the H$_2$ column densities. Our attempts to find a model satisfying the formula led to an unphysical model for the cloud in front of HD 37903. However we decided to present such model of the cloud towards HD 147888 for comparison purposes, as it has the hydrogen density at the observer side close to the density presented by \citet{Jensen}. 
The PDR model presented by \citet{Jensen} has n$_H$=2000 cm$^{-3}$. However, this model did not take into account the H$_2$ molecules in excited vibrational levels ($\nu>$0).

In the second model the hydrogen density varies from $n_H=18300$ cm$^{-3}$ near the observers side to $n_H=2604$ cm$^{-3}$ on the star side. The temperature changes from $T=34$ K to $T=131$ K, and the interstellar radiation field on the star side is 1 Draine unit. The star--cloud distance in the second model is 0.99 pc.
The ortho/para H$_2$ ratio in the cloud model equals to 1.4 and the total H$_2$ column density equals to 
$3.0\cdot10^{21}$ cm$^{-2}$. Column densities of H$_2$ energy levels in both models are shown on fig. \ref{lnN-E}.

\section{Conclusions}
   
  The ortho--para H$_2$ ratio on excited vibrational levels ($(O/P)_{\nu>0}=1.2$) differs from the O/P H$_2$ on the ground vibrational level ($(O/P)_{\nu=0,J=1}=0.16$).
  
  It is possible to obtain column densities from large number of shallow absorption lines arising from the same ground level. 
  The whole HST STIS spectrum should be fitted at once with a synthetic spectrum to derive column densities from a noisy spectrum.
 

\section*{Acknowledgments}

  This research was supported by University of Gda\'nsk grant BW/5400-5-0336-0.
  I would like to thank Herve Abgrall for providing the natural line widths for the H$_2$ lines.

\end{document}